\documentclass[prl,twocolumn,amssymb,showpacs,amsmath,nobibnotes,
superscriptaddress,floatfix,aps]{revtex4}
\usepackage{graphicx}
\usepackage{bm}
\usepackage{times}

\def\beq{\begin{equation}}
\def\eeq{\end{equation}}
\def\ba{\begin{array}}
\def\bea{\begin{eqnarray}}
\def\ea{\end{array}}
\def\eea{\end{eqnarray}}
\def\bit{\begin{itemize}}
\def\eit{\end{itemize}}
\def\nn{\nonumber}

\let\ad=\dagger

\let\ga=\gamma

\def\pbf{\bm{p}}

\def\xbf{\bm{x}}
\def\ybf{\bm{y}}

\def\to{\rightarrow}

\let\<=\langle
\let\>=\rangle

\begin{document}
\title{Semileptonic decays of $D$ mesons in three-flavor lattice QCD}
\author{C.~Aubin}
\author{C.~Bernard} 
\affiliation{Department of Physics, Washington University, St.~Louis, Missouri
63130}
\author{C.~DeTar} 
\affiliation{Physics Department, University of Utah, Salt Lake City, Utah 84112}
\author{M.~Di Pierro}
\affiliation{School of Computer Science, Telecommunications and Information
Systems, DePaul University, Chicago, Illinois 60604}
\author{A.~El-Khadra}
\affiliation{Physics Department, University of Illinois, Urbana, Illinois
61801-3080}
\author{Steven Gottlieb} 
\affiliation{Department of Physics, Indiana University, Bloomington, Indiana 47405}
\author{E.~B.~Gregory}
\affiliation{Department of Physics, University of Arizona, Tucson, Arizona 85721}
\author{U.~M.~Heller}
\affiliation{American Physical Society, One Research Road, Box 9000, 
Ridge, New York 11961-9000}
\author{J.~Hetrick}
\affiliation{University of the Pacific, Stockton, California 95211}
\author{A.~S.~Kronfeld}
\author{P.~B.~Mackenzie}
\affiliation{Fermi National Accelerator Laboratory, Batavia, Illinois 60510}
\author{D.~Menscher}
\affiliation{Physics Department, University of Illinois, Urbana, Illinois
61801-3080}
\author{M.~Nobes}
\affiliation{Physics Department, Simon Fraser University, Vancouver, British
Columbia, Canada}
\author{M.~Okamoto}
\affiliation{Fermi National Accelerator Laboratory, Batavia, Illinois 60510}
\author{M.~B.~Oktay}
\affiliation{Physics Department, University of Illinois, Urbana, Illinois
61801-3080}
\author{J.~Osborn}
\affiliation{Physics Department, University of Utah, Salt Lake City, Utah 84112}
\author{J.~Simone}
\affiliation{Fermi National Accelerator Laboratory, Batavia, Illinois 60510}
\author{R.~Sugar}
\affiliation{Department of Physics, University of California, Santa Barbara,
California 93106}
\author{D.~Toussaint}
\affiliation{Department of Physics, University of Arizona, Tucson, Arizona 85721}
\author{H.~D.~Trottier}
\affiliation{Physics Department, Simon Fraser University, Vancouver, British
Columbia, Canada}
\collaboration{Fermilab Lattice, MILC, and HPQCD Collaborations}
\noaffiliation

\date{August 26, 2004}
\pacs{11.15.Ha,12.38.Gc,13.20.Fc}
\begin{abstract}
We present the first three-flavor lattice QCD calculations for 
$D\to \pi l\nu$ and $D\to K l\nu$ semileptonic decays.
Simulations are carried out using ensembles of unquenched gauge fields
generated by the MILC collaboration.
With an improved staggered action for light quarks,
we are able to simulate at light quark masses down to 1/8 of the 
strange mass. Consequently, the systematic error from the chiral extrapolation
is much smaller than in previous calculations with Wilson-type light quarks.
Our results for the form factors at $q^2=0$ are 
$f_+^{D\to\pi}(0)=0.64(3)(6)$ and $f_+^{D\to K}(0) = 0.73(3)(7)$, where 
the first error is statistical and the second is systematic, added 
in quadrature. Combining our results with experimental branching ratios,
we obtain the CKM matrix elements
$|V_{cd}|=0.239(10)(24)(20)$ and $|V_{cs}|=0.969(39)(94)(24)$,
where the last errors are from experimental uncertainties.
\end{abstract}

\maketitle

Semileptonic decays of  heavy-light mesons are of great interest
because they can be used to determine CKM matrix elements such as
$|V_{ub}|$, $|V_{cb}|$, $|V_{cd}|$ and $|V_{cs}|$.
The accuracy of one of the most important, $|V_{ub}|$,
is currently limited by large theoretical uncertainty~\cite{Eidelman:wy}.
Lattice QCD provides a systematically improvable method of calculating
the relevant hadronic amplitudes, making the determination of $|V_{ub}|$
and other CKM matrix elements more reliable and precise.

Semileptonic $D$ meson decays, 
such as $D\to K l\nu$ and $D\to \pi l\nu$, provide a good test of
lattice calculations, because the corresponding CKM matrix elements
$|V_{cs}|$ and $|V_{cd}|$ are known more accurately than
$|V_{ub}|$~\cite{Eidelman:wy}.
The decay rates and distributions are not yet very well known, but the
CLEO-c experiment plans to measure them with an accuracy of a few per
cent.
Furthermore, measurements of leptonic and semileptonic $D_{(s)}$ decays 
can be combined so that the CKM matrix drops out, offering a direct and 
stringent check of lattice QCD.

Recently, dramatic progress has been achieved in lattice QCD, 
for a wide variety of hadronic quantities.
Reference~\cite{Davies:2003ik} showed agreement at the few per cent level
between three-flavor lattice QCD and experiment for $f_\pi$, $f_K$, 
mass splittings of quarkonia, and masses of heavy-light mesons.
The main characteristics of these quantities are that they have at most
one stable hadron in the initial and final states, and that the chiral 
extrapolation from simulated to physical light quark masses is under 
control.
This class can be called ``gold-plated''~\cite{Davies:2003ik}, and many
of the lattice calculations needed to test the Standard Model are in
this class.
The work reported here is part of a systematic effort to
calculate the hadronic matrix elements needed for leptonic and
semileptonic decays, and for neutral meson
mixing~\cite{diPierro:2003bu,Okamoto:2003ur}.

In this Letter we report results for $D\to K l\nu$
and $D\to \pi l\nu$ semileptonic decay amplitudes.
All previous lattice calculations of heavy-light semileptonic decays
have been done in quenched ($n_f=0$) QCD.
In addition to quenching, they also suffered from large uncertainties
from the chiral extrapolation and, in some cases, from large heavy-quark
discretization effects.
Here we bring all three uncertainties under good-to-excellent control.
Indeed, this Letter presents the first calculation in unquenched
three-flavor lattice QCD, where the effect of dynamical $u$, $d$ and $s$
quarks is correctly included.

The relevant hadronic amplitude $\< P | V^\mu | D \>~(P=\pi,K)$ is
conventionally parametrized by form factors $f_+$ and $f_0$ as
\bea
\< P | V^\mu | D \> =
f_+(q^2) (p_D+p_P-\Delta)^\mu + f_0(q^2) \Delta^\mu  
\eea
where $q = p_D - p_P$, 
$\Delta^\mu=(m_D^2-m_P^2)\, q^\mu / q^2$.
The differential decay rate $d\Gamma/dq^2$
is proportional to $|V_{cx}|^2 |f_+(q^2)|^2$, $x=d,s$.
(A contribution from $f_0$ is proportional to the lepton mass squared.)
We calculate $f_+$ and $f_0$ as a function of $q^2$ and determine
the decay rate $\Gamma$ and the CKM matrix
$|V_{cx}|$ by integrating $|f_+(q^2)|^2$
over $q^2$. Preliminary results have been reported in
Ref.~\cite{Okamoto:2003ur,Bernard:2003gu}.

Our calculations use ensembles of unquenched gauge fields generated by
the MILC collaboration~\cite{milc} with the ``Asqtad'' improved
staggered quark action and the Symanzik-improved gluon
action~\cite{asqtad}.
The results are obtained on the ``coarse'' ensembles
with sea quark masses 
$am_l^{\text{sea}}=0.005,0.007,0.01,0.02$, and 0.03.
The gauge coupling is adjusted to keep the same lattice cutoff 
($a^{-1}\approx 1.6~\text{GeV}$) and volume 
[$L^3\times T \approx (2.5~\text{fm})^3\times 8.0~\text{fm}$].
Each ensemble has about 400--500 configurations.
For details of the gauge configurations, see Ref.~\cite{milc}.

For the light valence quarks, we adopt the same 
staggered action as for the dynamical quarks.
The valence light ($u,d$) quark mass $m_l^{\rm val}$ is always set 
equal to $m_l^{\rm sea}$. 
The valence strange quark mass is 
$am_s^{\rm val}=0.0415$, 
which is slightly larger than the physical value $am_s=0.039$
(at this lattice spacing)
determined from fixing the masses
of the light pseudoscalars \cite{milc}.
We have repeated the calculations with a strange quark mass slightly too 
small, and find a negligible difference.
Since the computation of the staggered propagator is fast,
we can simulate with $m_l$ as low as~$m_s/8$.
Consequently we are able to reduce the systematic error from the chiral
extrapolation ($m_l\to m_{ud}$) to $\approx3\%$, as we show below.
In contrast, previous calculations with Wilson-type light quarks
simulated at $m_l\ge m_s/2$ and 
typically 
had
$O(10\%)$ errors 
from this source alone~\cite{El-Khadra:2001rv}. 

For the valence charmed quark we use the clover action 
with the Fermilab interpretation \cite{El-Khadra:1996mp}.
The bare mass is fixed via
the $D_s$ kinetic mass \cite{diPierro:2003bu}.
The free parameters of both the action and the current are adjusted so
that the leading heavy-quark discretization effects are
$O(\alpha_s a\Lambda_{\text{QCD}})$ and 
$O\left((a\Lambda_{\text{QCD}})^2\right)$, 
where $\Lambda_{\text{QCD}}$ is a measure of the QCD scale.

The hadronic matrix element $\< P | V^\mu | D \> $ is extracted from the
3-point function in the $D$ meson rest frame ($\pbf_D=0$)
\bea
C_{3,\mu}^{D\to P}(t_x,t_y; \pbf) =
\sum_{\xbf,\ybf} e^{i\pbf\cdot\ybf} \< O_P(0) \hat{V}_\mu(y) O_D^\ad(x) \> ,
\eea
where $\pbf=\pbf_P$,
$\hat{V}_\mu=\bar{\psi}_c\ga_\mu \psi_x$ ($x=d,s$)
is the heavy-light vector current on the lattice, and
$O_D$ and $O_P$ are interpolating operators for the initial and final 
states.
The heavy-light bilinears $\hat{V}_\mu$ and $O_D$ are formed from
staggered light quarks and Wilson heavy quarks as in
Ref.~\cite{Wingate:2002fh}.
The 3-point functions are computed  for 
light meson momentum $\pbf$ up to $2\pi(1,1,1)/L$,
using local sources and sinks.
The sink time is fixed typically to $t_x=20$.
To increase the statistics, the calculations are carried out 
not only at the source time $t_0=0$ but also at $t_0=16,32,48$ 
(and $t_x$ and $t_y$ shifted accordingly).
The results from 4 source times are averaged.
Statistical errors are estimated by the jackknife method.
To extract the transition amplitude $\<P|{V}^\mu|D\>$ we also need 
meson 2-point functions
$
C_2^M(t_x;\pbf) = \sum_{\xbf} e^{i\pbf\cdot\xbf} \<O_M(0)O_M^\ad(x)\>,
$
where $M=D,\pi,K$.
They are computed in an analogous way.
For the light meson ($M=\pi,K$) the 2-point function couples to the 
Goldstone channel of staggered quarks.

A drawback of staggered quarks is that each field produces 4~quark 
species, called ``tastes'' to stress that the extra 3 are unphysical.
One consequence is that the light meson 2-point function contains
a 16-fold replication of the desired hadrons.
On the other hand, the heavy-light 2-point function $C_2^{D}$ does
\emph{not} suffer from such replication, because contributions of heavy
quarks with momentum $p\sim O(\pi /a)$ are
suppressed~\cite{Wingate:2002fh}.
The same holds for 3-point functions that include at least one
Wilson quark, such as $C_{3,\mu}^{D\to P}$.
To check these properties, we carried out a preparatory quenched
calculation \cite{Okamoto:2003ur}, finding reasonable agreement with
those obtained previously with Wilson light
quarks~\cite{El-Khadra:2001rv}.

Another consequence of the extra tastes is that the 3- and 2-point 
functions
receive contributions from states that oscillate in 
time, in addition to the ground state and non-oscillating
excited state contributions.
For example, the 3-point function's time dependence takes the form
\bea
C_{3,\mu}^{D\to P}(t_x,t_y) &=& A_0 e^{-E_P t_y} e^{-E_D (t_x-t_y)}  \nn\\
&+& (-1)^{t_y} A_1 e^{-E' t_y} e^{-E_D (t_x-t_y)} + \cdots,\ 
\label{eq:C3}
\eea
where 
$A_0\propto\< P|V^\mu |D \>$.

As usual, the desired hadronic amplitude is extracted from fitting the 3- 
and 2-point functions.
We employ two methods.
In the first method, we form the ratio
$R(t_y)\equiv C_{3,\mu}^{D\to P}(t_x,t_y)/[C_2^{P}(t_y) C_2^{D}(t_x-t_y)]$, 
and fit to a constant in $t_y$.
The oscillating state contributions are partly canceled in the ratio, and further
reduced by taking the average, $\tilde{R}(t_y)=[R(t_y)+R(t_y+1)]/2$.
A plateau is then found for $t_y$ around $t_x/2$.
In the second method, we first fit $C_{3,\mu}^{D\to P}$ and $C_2^{P,D}$
separately, using a multi-exponential form similar to Eq.~(\ref{eq:C3}),
and then obtain $\< P|V^\mu |D \>$ from the fit results.
The results from the two methods always agree within statistical errors.
The difference between two results  
is less than $3\%$ for the lower two momenta, and as large as $3\%$ for 
the higher two momenta.
We choose 
the first method for central values and take 3\%
as the systematic error from the fitting.

The lattice heavy-light vector current must be multiplied by a
renormalization factor $Z_{V_\mu}^{cx}$.
We follow the method in Ref.~\cite{El-Khadra:2001rv}, writing
$Z_{V_\mu}^{cx}=\rho_{V_\mu}(Z_{V}^{cc} Z_{V}^{xx})^{1/2}$.
The flavor-conserving renormalization factors $Z_{V}^{cc}$ and 
$Z_{V}^{xx}$ are computed nonperturbatively
from standard charge normalization conditions.
The remaining factor $\rho_{V_\mu}$
is expected to be close to unity because most of the radiative
corrections are canceled in the ratio~\cite{Harada:2001fi}.
A one-loop calculation
gives \cite{one-loop} $\rho_{V_4} \approx 1.01$ and $\rho_{V_i} \approx 0.99$
which we use in the analysis below.
This perturbative calculation is preliminary, but it has been 
subjected to several non-trivial tests.

Rather than calculating the conventional form factors $f_0$ and $f_+$ directly,
we first extract 
the 
form factors $f_\parallel$ and $f_\perp$, 
as in Ref.~\cite{El-Khadra:2001rv}, defined through
\bea
\< P | V^\mu | D \>
&=& 
\sqrt{2m_D} \, \left[v^\mu \, f_\parallel(E) +
p^\mu_\perp \, f_\perp(E) \right], \nonumber
\eea
where $v=p_D/m_D$, $p_\perp=p_P-Ev$ and $E=v\cdot p_P$ 
is the energy of the light meson.
$f_\parallel$ and $f_\perp$ are more natural quantities in the heavy quark 
effective theory, and chiral expansions are given for them 
as a function of $E$ \cite{Becirevic:2002sc,SchiPTheavy}.
We therefore carry out the chiral extrapolation in $m_l$ 
for $f_\parallel$ and $f_\perp$ at fixed $E$, and then 
convert to $f_0$ and $f_+$.

\begin{figure}[t]
\includegraphics*[width=5.9cm]{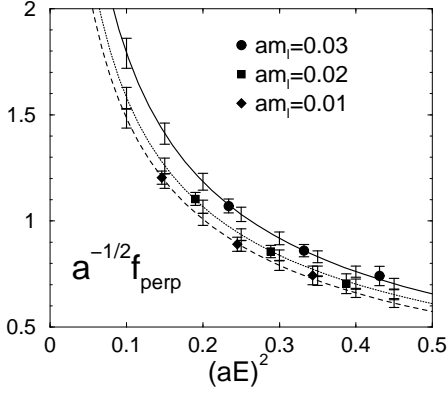}
\caption{$a^{-1/2}f_\perp$ as a function of $(aE)^2$ for the $D\to\pi$ decay. 
Symbols are raw data and 
lines are fitting curves with the parametrization of Eq.~(\ref{eq:BK}).
Results at $m_l=0.03,0.02$ and 0.01 are shown.
}
\label{fig:pdep}
\end{figure}

To perform the chiral extrapolation at fixed $E$, we interpolate and
extrapolate the results for $f_\parallel$ and $f_\perp$ to common values
of $E$.
To this end, we fit $f_\parallel$ and $f_\perp$ simultaneously using
the parametrization of Becirevic and Kaidalov (BK)
\cite{Becirevic:1999kt},
\begin{equation}\label{eq:BK}
f_+(q^2) = \frac{F}{(1-\tilde{q}^2)(1-\alpha\tilde{q}^2)},~~~
f_0(q^2) = \frac{F}{1-\tilde{q}^2/\beta},
\end{equation}
where $\tilde{q}^2=q^2/m_{D_x^{*}}^2$, and $F=f_+(0)$, $\alpha$ and $\beta$
are fit parameters, and $f_+$, $f_0$ and $q^2$ are converted to
$f_\parallel$, $f_\perp$ and $E$ before the fits.
An advantage of the BK form is that it contains a pole in $f_+(q^2)$ 
at $q^2=m^2_{D^*_x}$, where $m_{D^*_x}$ is the lattice mass of
the charmed vector meson with daughter quark $x$.
The BK fit for $f_{\perp}$ is shown in Fig.~\ref{fig:pdep},
using data for all available momenta~$\pbf$.
Excluding the data for the highest momentum $2\pi(1,1,1)/L$ gives
indistinguishable results.

\begin{figure}[t]
	\includegraphics*[width=5.9cm]{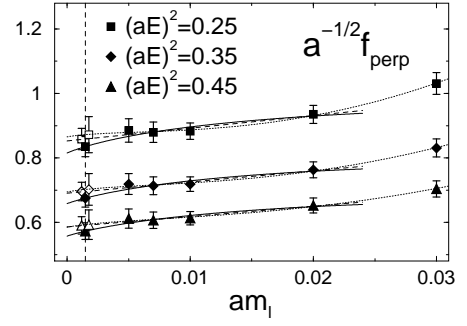}
	\caption{
        $m_l$-dependence and chiral fits for $a^{-1/2}f_\perp^{D\to\pi}$ for 
	several values of $(aE)^2$. The S$\chi$PT+linear fit(solid),
	S$\chi$PT+quadratic fit(dotted) and linear fit(dashed).}
	\label{fig:chiral}
\end{figure}

We perform the chiral extrapolation using recently obtained 
expressions \cite{SchiPTheavy} for heavy-to-light form factors  in
staggered chiral perturbation theory (S$\chi$PT) \cite{Bernard:2001yj}.
Compared with the continuum $\chi$PT formulae \cite{Becirevic:2002sc}, 
the S$\chi$PT formulae include 
6 new parameters (4 splittings and 2 taste-violating hairpins),
which parameterize lattice discretization effects.
The new parameters are fixed
from the analysis of light pseudoscalars \cite{milc}.
As usual, low energy constants appear, such as the chiral coupling $f$
and heavy-to-light meson coupling $g$.
We take $f=130$ MeV and $g=0.59$, but changing these constants by 10\%
has negligible effect.
The fit form we adopt (``S$\chi$PT+linear'') is
\beq\label{eq:schiPT}
f_{\perp,\parallel}(E) = A [1+\delta f_{\perp,\parallel}(E)] + B m_l,
\eeq
where $A,B$ are fit parameters, 
and $\delta f_{\perp,\parallel}$ is the S$\chi$PT correction. 
To estimate the systematic error here, we try a 
simple linear fit and a ``S$\chi$PT+quadratic'' fit with a term $C m_l^2$ 
added to Eq.~(\ref{eq:schiPT}).
A comparison of the three fits is shown in Fig.~\ref{fig:chiral}. 
For the $D\to\pi~(K)$ decay
the linear fit gives 3\% (2\%) larger results
at $m_l=m_{ud}$.
The results from the S$\chi$PT+quadratic fit typically lie 
between the results from the other two fits, with larger errors. 
We therefore take 3\% (2\%) as the systematic error from the chiral 
extrapolation for the $D\to\pi~(K)$ decay.

\begin{figure}[t]
\includegraphics*[width=5.9cm]{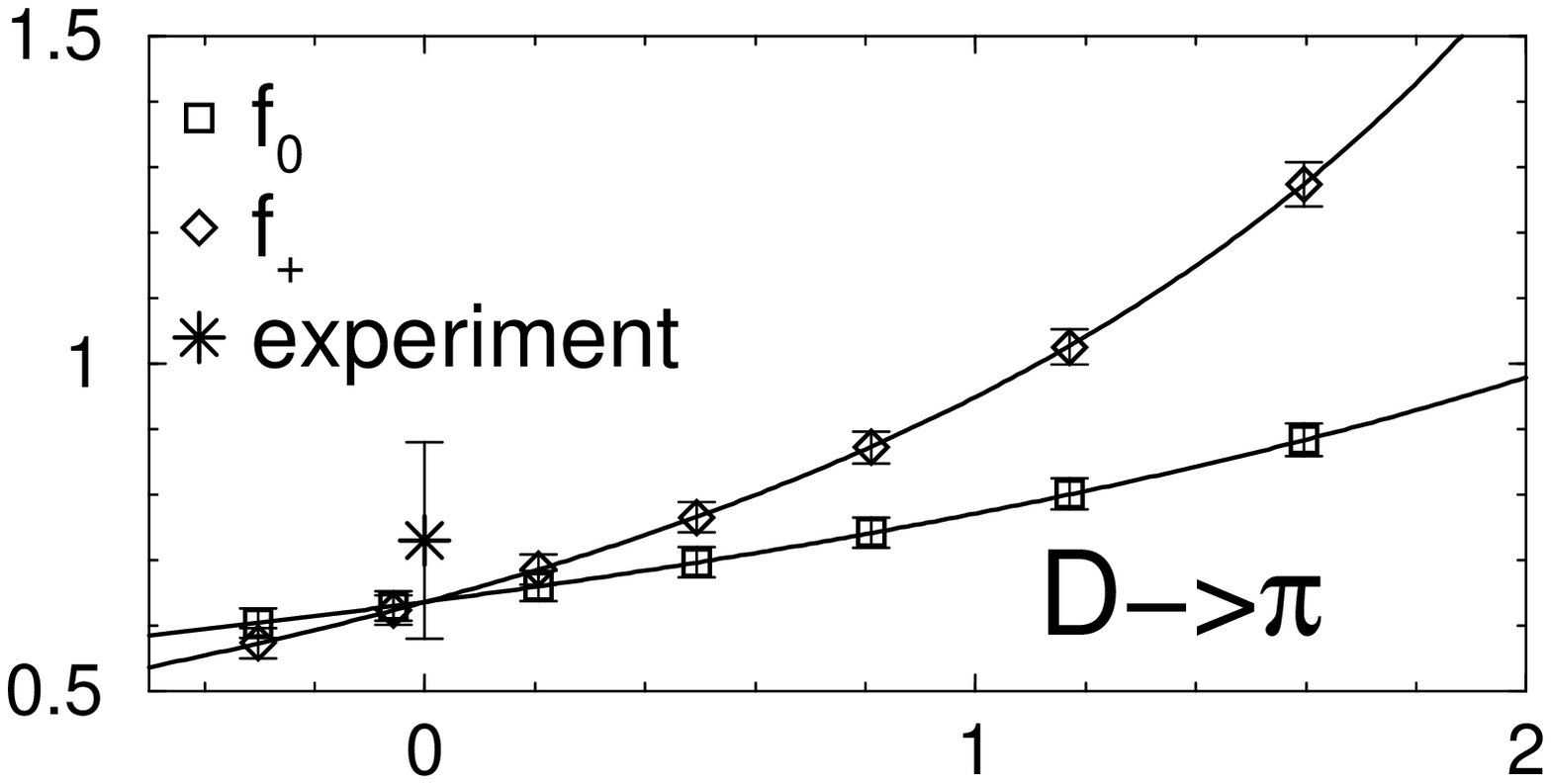}
\includegraphics*[width=5.9cm]{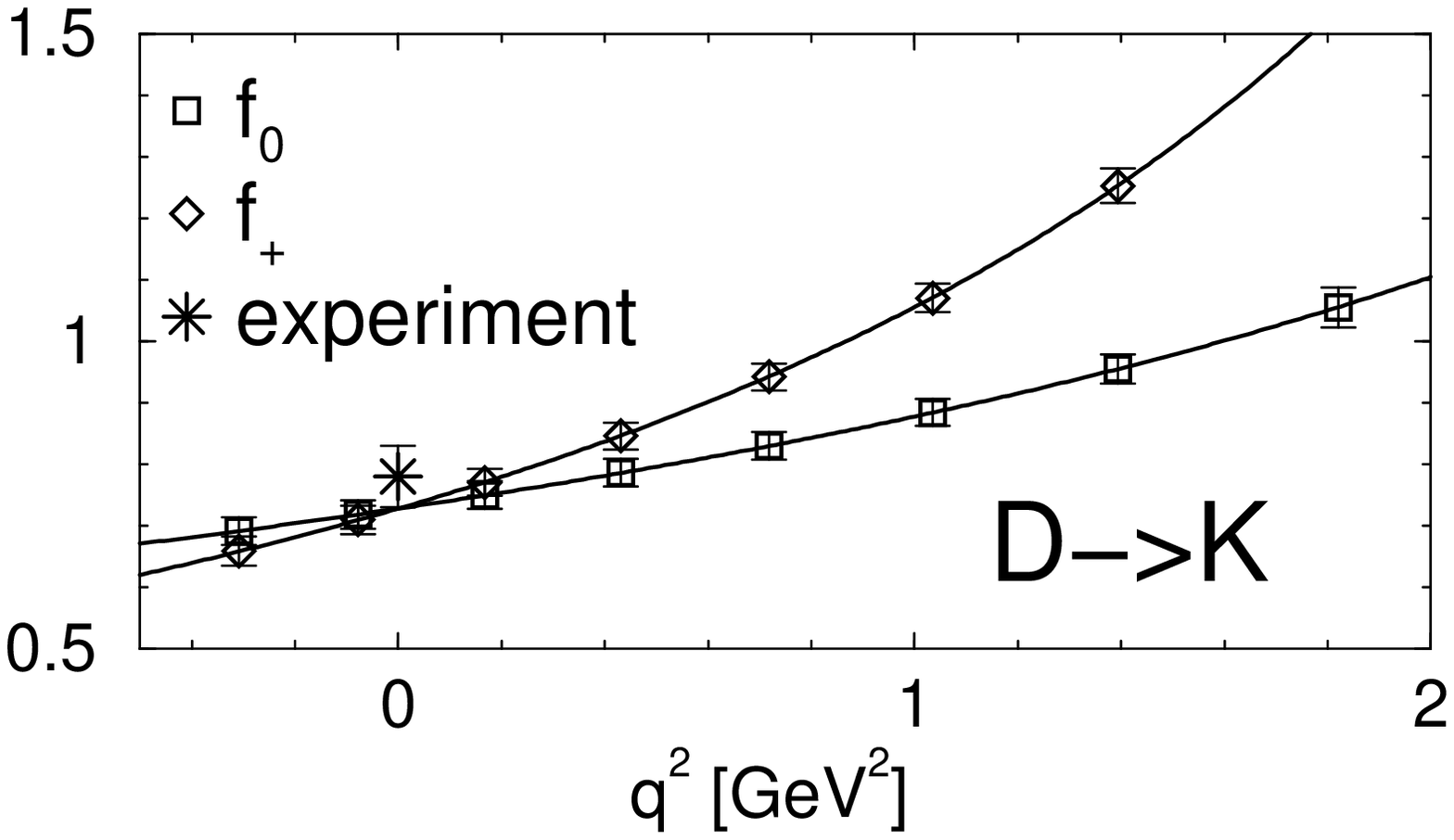}
\caption{$D\to\pi$ and $D\to K$ form factors.
The experimental values are taken from Ref.~\cite{unknown:2004nn}.}
\label{fig:D2piK}
\end{figure}

\begin{table}[t]
\caption{Fit parameters in Eq.~(\ref{eq:BK}), decay rates and 
CKM matrix elements.
The first errors are statistical; the second systematic;
the third experimental.}
\label{tab:result}
\begin{tabular}{cccccc}
\hline\hline
$P$ & $F$& $\alpha$ & $\beta$ & $\Gamma/|V_{cx}|^2~[\text{ps}^{-1}]$
& $|V_{cx}|$\\ \hline
$\pi$ &0.64(3) &0.44(4) &1.41(6)  &0.154(12)(31) & 0.239(10)(24)(20)\\
 $K$  &0.73(3) &0.50(4) &1.31(7)  &0.093(07)(18) & 0.969(39)(94)(24)\\
\hline\hline
\end{tabular}
\end{table}

We now convert the results for $f_{\perp}$ and $f_{\parallel}$ at $m_l=m_{ud}$,
to $f_+$ and $f_0$.
To extend  
$f_+$ and $f_0$ to functions of $q^2$, 
we again fit to the form Eq.~(\ref{eq:BK}).
The results are shown in Fig.~\ref{fig:D2piK}, with statistical errors only.
We then obtain the decay rates $\Gamma/|V_{cx}|^2$ by integrating 
$(\text{phase space})\times|f_+(q^2)|^2$ over $q^2$.
Finally, we determine the CKM matrix elements $|V_{cd}|$ and $|V_{cs}|$
using experimental lifetimes and branching ratios~\cite{Eidelman:wy}.
These main results are summarized in Table~\ref{tab:result}. 

The results presented above 
rely on the $q^2$ dependence of BK parametrization, Eq.~(\ref{eq:BK}).
To estimate the associated systematic error, we 
make an alternative analysis without it.
We perform a 2-dimensional fit in $(m_l,E)$ to the raw data employing 
a polynomial form plus the S$\chi$PT correction $\delta 
f_{\parallel,\perp}$.
The result from this fit agrees with 
the
one from the fit with Eq.~(\ref{eq:schiPT})
within statistical errors. The deviation between the two fits is 
negligible at $q^2 \sim q^2_{\rm max}$ and about 1$\sigma$ at $q^2 \sim 0$
for $f_{\perp,\parallel}$, 
giving a 2\% difference for the CKM
matrix elements.

With only one lattice spacing, the systematic error from discretization
effects can be estimated only by power counting.
The leading discretization errors from the Asqtad action are
$O\left(\alpha_s (a\Lambda_{\text{QCD}})^2\right)\approx 2\%$ 
(after removal of taste-violating effects with S$\chi$PT), taking
$\Lambda_{\text{QCD}}=400$~MeV and $\alpha_s=0.25$. 
In addition, there is a momentum-dependent error from the final state.
The BK parameters are determined by the lower momentum data;
in particular, the fits are insensitive to the highest
momentum $2\pi(1,1,1)/L$.
Therefore we estimate this effect to be
$O(\alpha_s (a\pbf)^2)\approx 5\%$, taking the second-highest
momentum $\pbf=2\pi(1,1,0)/L$.
The HQET theory of cutoff effects~\cite{Kronfeld:2000ck,Kronfeld:2003sd}
can be used to estimate the discretization
error from the heavy charmed quark.
In this way, we estimate the discretization error to be 4--7\%,
depending on the value chosen for $\Lambda_{\text{QCD}}$ (in the HQET 
context).
This is consistent with the lattice spacing dependence seen in
Ref.~\cite{El-Khadra:2001rv}.
In future work we expect to reduce and understand better this 
uncertainty, so we shall adopt the maximum, 7\%, here.

A summary of the systematic errors for the form factors
$f_{+,0}$ or the CKM matrix elements $|V_{cx}|$ is
as follows.
The error from time fits is 3\%; from chiral fits, 
3\% (2\%) for $D\to\pi~(K)$ decay; 
from BK parametrization, 2\%.
The 1-loop correction
to $\rho_{V_\mu}$ is only 1\%, so 2-loop uncertainty is 
assumed to be
negligible.
The uncertainty for $a^{-1}$ is about 1.2\% \cite{milc};
this leads to a 1\% error for $|V_{cx}|$ (but not for the dimensionless 
form factors), from integrating over $q^2$ to get $\Gamma/|V_{cx}|^2$.
Finally, we quote discretization uncertainties of
2\%, 5\%, and 7\%, from light quarks, the final state energy, and the 
charmed quark, respectively.
Adding all the systematic errors in quadrature, we find the total 
to be $[3\% + 3\%~(2\%) + 2\% + 1\% + 2\% + 5\% + 7\%] = 10\%$.

Incorporating the systematic uncertainties, we obtain
\begin{eqnarray}
    f_+^{D\to\pi}(0) & = & 0.64(3)(6), \\
    f_+^{D\to K} (0) & = & 0.73(3)(7),
\end{eqnarray}
and the ratio $f_+^{D\to\pi}(0)/f_+^{D\to K}(0)=0.87(3)(9)$.
Our results for the CKM matrix elements (Table~\ref{tab:result})
are consistent with Particle Data Group averages
$|V_{cd}|=0.224(12)$ and $|V_{cs}|=0.996(13)$~\cite{Eidelman:wy};
also with $|V_{cs}|=0.9745(8)$ from CKM unitarity. 
If we instead use these CKM values as inputs, we 
obtain,
for the total decay rates,
\bea
\Gamma(D^0\to\pi^- l^+\nu) & \!=\! &
    (7.7 \pm 0.6 \pm 1.5 \pm 0.8)\times 10^{-3} {\rm ps}^{-1},\nn\\
\Gamma(D^0\to K^-  l^+ \nu)& \!=\! &
    (9.2 \pm 0.7 \pm 1.8 \pm 0.2)\times 10^{-2} {\rm ps}^{-1},\nn\\
\frac{\Gamma(D^0\to\pi^- l^+\nu)}{\Gamma(D^0\to K^-  l^+ \nu)} & \!=\! &
    0.084 \pm 0.007 \pm 0.017 \pm 0.009,
\label{eq:rate}
\eea
where the first errors are statistical, the second systematic, and
the third from uncertainties in the CKM matrix elements. 
We do not assume any cancellation of errors in the ratios,
although some may be expected.
Our results agree with recent experimental results,
$f_+^{D\to\pi}(0) = 0.73(15)$, $f_+^{D\to K}(0) = 0.78(5)$
\cite{unknown:2004nn},
$f_+^{D\to\pi}(0)/f_+^{D\to K}(0) =  0.86(9)$
and 
$
\Gamma(D^0\to\pi^- e^+\nu_e)/\Gamma(D^0\to K^-  e^+ \nu_e)
= 0.082 \pm 0.008
$~\cite{Huang:2004fr}. 

This Letter presents the first three-flavor lattice calculations
for semileptonic $D$ decays.
With an improved staggered light quark,
we have successfully reduced the two dominant uncertainties of
previous works, \emph{i.e.}, the effect of the quenched 
approximation and the error from chiral extrapolation.
Our results for the form factors, decay rates and CKM matrix,
given in Table~\ref{tab:result} and Eq.~(\ref{eq:rate})
are in agreement with experimental results.
The total size of systematic uncertainty is 10\%, which is dominated by 
the discretization errors.
To reduce this error,
calculations at finer lattice spacings
and with more highly-improved heavy-quark actions are 
necessary; these are underway.
Finally,
unquenched calculations of 
$B$ decays such as $B\to\pi l\nu$ and 
$B\to D l\nu$ are 
in progress, 
and will be presented in a separate paper. 

We thank the Fermilab Computing Division, the SciDAC Program and the
Theoretical
High Energy Physics Programs at the DOE and NSF for their support.
Fermilab is operated by Universities Research Association Inc., under
contract with the U.S. Department of Energy.

\end{document}